\documentclass[12pt]{article}
\usepackage[pctex32]{graphics}
\begin{document}
\vspace{1cm}
\begin{center}
~\\
{\bf  \Large  Quantum Field, Thermodynamics and Black Hole on Coherent State Representation of Fuzzy Space}
\vspace{1cm}

                      Wung-Hong Huang\\
                       Department of Physics\\
                       National Cheng Kung University\\
                       Tainan, Taiwan\\

\end{center}
\vspace{1cm}
\begin{center}{\bf  \Large ABSTRACT } \end{center}
We first use the coherent state formalism of fuzzy space to show that the  fuzziness  will eliminate point-like structure of a particle in favor of smeared object, which is an exponential decay function in contrast to the Gaussian type in the Moyal noncommutative space.   The exponential decay function implies that, in the UV region, the fuzziness provides an extra power-decay factor in the Feynman propagator, contrasts to the exponential-decay factor in the Moyal space.  We also calculate the particle heat capacity and see that it approaches to zero at high temperature.  Next, we use the found smeared source to study the  Schwarzschild-like geometry and see that the black hole can reach a finite maximum temperature before cooling down to absolute zero and leave a stable remnant, as that in the noncommutative case.  The properties of fuzzy 3D BTZ and the fuzzy Kaluza-Klein black holes are also discussed.   Finally, we present a criterion for existence a regular black hole with a general smeared source function.
\vspace{4cm}
\\
\begin{flushleft}
*E-mail:  whhwung@mail.ncku.edu.tw\\
\end{flushleft}
\newpage
\section{Introduction}
Physics on the noncommutative spacetime had been received a great deal of attention [1-3].  Historically, it is a hope that the deformed geometry in the small spacetime would be possible to cure the quantum-field divergences, especially in the gravity theory.  The renovation of the interesting in noncommutative field theories is that it has proved to arise naturally in the string/M theories [4,5].  The noncommutative geometry is encoded in the commutator 
$$[ x_\mu, x_\nu] = i \theta_{\mu\nu},\eqno{(1.1)} $$
The value of $\theta_{\mu\nu}$ is an anti-symmetric matrix which measures the space noncommutativity.   Relation (1.1) implies the $\star$ operator which is the Moyal product generally defined by
$$\Phi(x) \star \Phi(x) = e^{+{i\over 2} \theta^{\mu\nu} {\partial\over \partial
y^\mu} {\partial\over \partial z^ \nu} } \Phi(y) \Phi(z) |_{y,z\rightarrow x}.  
\eqno{(1.2)} $$
Using the Moyal product the quantum field on the noncommutative geometry had been studied extensively [1-5], including the thermal property [6].

 There is another interesting approach to the noncommutative quantum theory : the coordinate coherent state approach [7,8]. Let us first briefly review the approach.

  Consider coherent state on the 2D noncommutative space in which the noncommutative coordinates $\hat q_1$ and $\hat q_2$ have the property
$$[\hat q_1,\hat q_2] = i\theta.\eqno{(1.3)}$$
 We first perform the canonical transformation by defining the operators 
$$\hat A= \hat q_1+ i \hat q_2,~~~~~\hat A^\dag = \hat q_1- i \hat q_2, ~~~~~\Rightarrow~~~~~~ [\hat A,\hat A^\dag]= 2\theta. \eqno{(1.4)}$$
The coherent state $ |\alpha \rangle$ is defined by  
$$ \hat A|\alpha\rangle = \alpha|\alpha\rangle ,~~ \langle\alpha|\hat A^\dag  = \langle\alpha|\alpha^* ~~~~\Rightarrow~~~|\alpha\rangle = e^{-{|\alpha|^2\over2}}e^{\alpha A^\dag}.   \eqno{(1.5)}$$
The  coordinates  $x_1$, $x_2$ which represent the mean position of the particle over the non-commutative plane are defined by
 $$x_1 \equiv \langle\alpha|\hat q_1|\alpha \rangle=  \langle\alpha| {\hat A + \hat A^\dag} |\alpha \rangle={\alpha + \alpha^*}.   \eqno{(1.6)}$$
 $$x_2 \equiv \langle\alpha|\hat q_2|\alpha \rangle=  \langle\alpha| {\hat A - \hat A^\dagger} |\alpha \rangle={\alpha- \alpha^*}.  \eqno{(1.7)}$$
The noncommutative version of the plane wave operator is defined by $\exp i\left(\vec p\cdot   \vec {\hat q}\right)$.  Using the Baker-Campbell-Hausdorff  formula the mean value becomes
 $$\langle \alpha| e^{ i p_1\hat q_1 +i p_2 \hat q_2}|\alpha \rangle =
  \langle \alpha|e^{i (p_+ c_+ +p_- c_-)\hat A^\dag +i (p_+ c_- +p_- c_+)\hat A}|\alpha \rangle = e^{\left(ip_1x_1+ip_2 x_2 -{\theta\over4} ( p_1^2+ p_2^2)\right)}.\hspace{1.3cm}\eqno{(1.8)}$$
In the above calculation we have defined $ p_\pm\equiv ( p_1 \pm i\, p_2)/2$.  According to the [7,8] we can interpret (1.8)  as the wave function of a ``free point particle'' on the non-commutative plane:
$$ \Psi_{\vec p}(\vec x) =\langle\vec p|\vec x\rangle =exp\left(-\theta {\left(p_1^2 
+ p_2^2 \right)\over4}+i \vec p\cdot \vec x\right). \eqno{(1.9)}$$
After the Fourier transformation of above result the mass density of the point particle become a smeared one described by
$$\rho_\theta= {M\over 2\pi\theta}exp\left(-{\vec x ^2\over2\theta}\right), \eqno{(1.10)}$$
which tells us that the effect of non-commutativity in the scalar product between two mean positions is to substitute a Dirac delta function by a Gaussian distribution with  half-width $\theta$  and we have a smeared source now.

  Many authors had used the above smeared source  to study the effects of noncommutativity on the quantum field [7,8] and terminal phase of black hole evaporation [9-13].  Some interesting results have found  are : there exists a finite maximum temperature that the black hole can reach before cooling down to absolute zero;  there is no curvature singularity at the origin while existence a regular De-Sitter core at short distance.  

There is another noncommutative geometry, called as fuzzy space, which appear naturally in the string/M theory [14,15] too.  It is known to correspond to the sphere D2-branes in string theory with background linear B-field. Also, in the presence of constant RR field potential, the D0-branes can expand into a noncommutative fuzzy sphere configuration.  The quantum field on the fuzzy space had been studied by many authors [16-18].   Therefore, it is interesting to see how the fuzzy space will modify the black hole property and in this paper we will use the coherent state approach to study the problem.  

Note that  in Moyal noncommutative geometry in which the commutation of  coordinates $x_i$ is a constant value, as shown in (1.1).  Therefore, using the Baker-Campbell-Hausdorff  formula we can find a  simple relation  (1.8). However, on the fuzzy sphere with coordinates $X_i$, which is now described by $ [X_i, X_j]= i h X_k$.  Thus the commutation of  coordinates $X_i$ is not a constant value and the Baker-Campbell-Hausdorff  formula could not  lead to a simple result. In section II we will use an approximation to study the associated smeared source.   After a lengthy calculation we in section III find that the smeared object is an exponential decay function in contrast to the Gaussian type in the Moyal noncommutative space.

   In section IV we  see that the exponential decay function could, in the UV region, provides Feynman propagator with an extra power-decay factor $(hk)^{-4}$, contrasts to the exponential-decay factor $e^{-\theta k^2}$  in the Moyal space.   We also calculate the particle heat capacity and see that it approaches to zero at high temperature.

 Using the found smeared source we then in section V study the associated black hole property and compare it with that on the Moyal noncommutative geometry.  Especially, in section VI we present a criterion for existence a regular black hole with a general smeared source.  Last section is devoted to a short discussion.
\section {Coherence State Representation of Fuzzy Space}
\subsection {Fuzzy Space  $R_{ h }^3$ and Coherence State $|w\rangle$}
It is known  that a collection of N D0-branes in a constant background RR  field could expand as fuzzy sphere described by the coordinates
$$ [X_i, X_j] = i h ~\epsilon_{ijk} X_k,~~~~~i,j,k=1,2,3,     \eqno{(2.1)}$$
in which $ h $  is proportional to the strength of RR field.   A non-trivial solution is $X_i={ h \over2}  J_i$ in which $J_i$ is the $J\times J$ matrix representation of the SU(2) algebra.  It describe a fuzzy sphere, $S_ h ^2$.   In other words the fuzzy sphere $S_{ h ,J}^2$  is determined by the algebra (2.1) subjected to 
$$ \sum _{i=1}^{3} X_i^2 =  h ^2 J(J+1) = r^2,     \eqno{(2.2)}$$
and the radius of the fuzzy sphere is given by $r= h  \sqrt{ J(J+1)}$. Therefore the full noncommutative 3D space $R_{ h }^3$ can be obtained when we consider the set of fuzzy spheres with all possible radii, i.e.
$$R_{ h }^3 = \sum_{J=0}^\infty S_{ h ,J}^2. \eqno{(2.3)}$$
Notice that $R_{ h }^3$ does not reduce to commutative $R^3$ merely letting $ h  \rightarrow 0$ [17].  Only in the limit $J\rightarrow \infty$ while at fixed $R$ can we get the ordinary sphere with radius $R$. 

   The coherence state on fuzzy space is given by [19]
$$ |w\rangle = {1\over (1+|w|^2)^{2J}} e^{w X_+/ h } |J,-J\rangle = {1\over (1+|w|^2)^{2J}}\sum_{n=0}^{2J} w^n\left({2J! \over (2J-n)!n!}\right)|J,-J+n\rangle.\eqno{(2.4)}$$
The uncertainty relation for the $J_i$ operators defined in the algebra (2.1) is given by 
$$\langle X_i^2\rangle \langle X_j^2\rangle \ge { h ^2\over 4}\langle X_i^2\rangle^2.\eqno{(2.5)}$$
This equality holds for the state $|w\rangle$, in which $\langle X_i^2\rangle \equiv \langle w| X_i^2|w\rangle$ ; hence the coherent states are the minimum-uncertainty states [19].
\subsection{Operators $\hat A$ and $\hat B$ }
We now begin our calculations.  First, it is easy to check that the coherent state  satisfies the following relations
$$ X_- |w\rangle =  w( h  J-X_z)|w\rangle.\eqno{(2.6)}$$
$$ X_+ |w\rangle =  {1\over w}( h  J+X_z)|w\rangle.\eqno{(2.7)}$$
Therefore we have the following two important operators $\hat A$ and $\hat B$ which have coherent state as their eigenstate, i.e. 
$$ \hat A \equiv (X_-+wX _z)/ h ,~~~~~\Rightarrow~~~\hat A|w\rangle =  wJ|w\rangle,~~~~~\langle w| \hat A^\dag = \langle w|w^*J.\eqno{(2.8)}$$
$$ \hat B \equiv (X_+ - {1\over w}X _z)/ h , ~~~~~\Rightarrow~~~\hat B|w\rangle = {1\over w} J|w\rangle~~~~~~\langle w| \hat B^\dag= \langle w| {1\over w^*} J.\eqno{(2.9)}$$
Above definition implies following relations
$$X_+ =  {h\over 1+|w|^2}(\hat A^\dag +|w|^2\hat B).\hspace{3cm}\eqno{(2.10)}$$
$$X_- =  {h\over 1+|w|^2}(\hat A+\hat |w|^2 \hat B^\dag).\hspace{3cm}\eqno{(2.11)}$$
$$X_z =  {h w^*\over 1+|w|^2}(\hat A-\hat B^\dag) = {h w\over 1+|w|^2}(\hat A^\dag-\hat B).\eqno{(2.12)}$$
The operators $\hat A$ and $\hat B$  defined in (2.8) and (2.9) satisfy the following  commutation relations.
$$[ h  \hat A^\dag,  h  \hat A] = - h^2w^* \hat B^\dag - h^2w\hat B. \eqno{(2.13)}$$
$$[ h  \hat A^\dag,  h  \hat B^\dag] = {h^2\over w^*} \hat A^\dag - h^2w^* \hat B^\dag. \eqno{(2.14)}$$
$$[ h  \hat A^\dag,  h  \hat B] = {h^2\over w} \hat A^\dag +h^2w^* \hat B  . \eqno{(2.15)}$$
$$[ h  \hat A,  h  \hat B] = - {h^2\over w} \hat A + h^2w \hat B. \eqno{(2.16)}$$
$$[ h  \hat B^\dag,  h  \hat B] =  - {h^2\over w^*}\hat A^\dag-{h^2\over w} \hat A. \eqno{(2.17)}$$
Using (2.8) and (2.9) above relations (2.13)-(2.17) imply that
$$\langle w| exp( c [ h  \hat A^\dag,  h  \hat A]|w\rangle = exp(  -2 c~h^2J). \eqno{(2.18)}$$
$$\langle w| exp( c [ h  \hat A^\dag,  h  \hat B^\dag]|w\rangle = exp( 0)=1. \eqno{(2.19)}$$
$$\langle w| exp( c [ h  \hat A^\dag,  h  \hat B]|w\rangle = exp( 2ch^2{w^*\over w} J). \eqno{(2.20)}$$
$$\langle w| exp( c [ h  \hat A,  h  \hat B]|w\rangle = exp( 0)=1. \eqno{(2.21)}$$
$$\langle w| exp( c [ h  \hat B^\dag,  h  \hat B]|w\rangle = exp(-2c~h^2J), \eqno{(2.22)}$$
to the leading order of $h^2$.  The property will be used in next subsection.
\subsection {Plane Wave on Fuzzy Space}
To proceed, we first express the mean value of plane wave as the following form : 
 $$\langle w| exp( i p_-X_++i p_+X_- +i p_z X_z)|w \rangle =  \langle w|exp\left({i\over 2(1+|W|^2)}\right)\cdot\hspace{6cm}$$
$$\left( (p_-+p_z w) h \hat A^\dag + (p_+|w|^2- p_z w^*)  h \hat B^\dag+ (p_++p_z w^*) h \hat A + (p_-|w|^2- p_z w)  h \hat B \right)|w \rangle.\eqno{(2.23)}$$
Here we have used the definitions (2.10)-(2.12). 

Next, from (2.13)-(2.17) we  see that the commutations between operators $ h \hat A^\dag$, $ h \hat A$, $ h \hat B^\dag$ and $ h \hat B$ are proportional to $h^2$.   Therefore we can use the formula
$$e^{M_1+M_2+M_3+M_4}\approx e^{M_1} e^{M_2} e^{-{1\over2}[M_1,M_2]}  e^{M_3} e^{M_4}  e^{-{1\over2}[M_3,M_4]}e^{-{1\over2}[M_1,M_3]}e^{-{1\over2}[M_1,M_4]}e^{-{1\over2}[M_2,M_3]}e^{-{1\over2}[M_2,M_4]}.\eqno{(2.24)}$$
in which $M_i$ are matrix operators, to evaluate (2.23) to the leading order of  $ h $, with the help of commutative relations (2.18)-(2.22).  

Now, identify
$$M_1= exp\left({i\over 2(1+|W|^2)}~(p_-+p_z w) h \hat A^\dag \right).\eqno{(2.25)}$$
$$M_2= exp\left({i\over 2(1+|W|^2)}~(p_+|w|^2- p_z w^*)  h \hat B^\dag \right).\eqno{(2.26)}$$
$$M_3= exp\left({i\over 2(1+|W|^2)}~(p_++p_z w^*) h \hat A \right).\eqno{(2.27)}$$
$$M_4= exp\left({i\over 2(1+|W|^2)}~(p_-|w|^2- p_z w)  h \hat B  \right),\eqno{(2.28)}$$
as those in (2.24), using (2.18)-(2.22) and after a lengthy calculation we find that the plane wave on fuzzy space could be expressed as an elegant result 
 $$\langle\vec p|\vec x\rangle=\langle w| exp( i p_-X_++i p_+X_- +i p_z X_z)|w \rangle = exp\left( i \vec x \cdot \vec p -{|\vec x|\over 4  h}(\vec x^2  \vec p^2 - (\vec x \cdot \vec p)^2)\right).\eqno{(2.29)}$$
This relation corresponds to the smeared source on Moyal space described in (1.9).  The classical value $\vec x$ in (2.29) is the mean value of  (2.10)-(2.12) in the coherent state $|w\rangle$, i.e.
$$x_x+ix_y\equiv x_+ \equiv \langle w|X_+|w\rangle = {2hw^*J\over 1+|w|^2}.\eqno{(2.30)}$$
$$x_x-ix_y\equiv x_- \equiv \langle w|X_-|w\rangle = {2hwJ\over 1+|w|^2}.\eqno{(2.31)}$$
$$x_z\equiv \langle w|X_z |w\rangle= {(|w|^2-1)hJ\over 1+|w|^2},\hspace{1cm}\eqno{(2.32)}$$
after using the properties of (2.8) and (2.9).
\section {Smeared Source on Fuzzy Space}
As that in Moyal space, after taking the Fourier transformation of plane wave (2.29) the mass density of the point particle will become a smeared one.   However, the mathematic difficulty in there  forbid us to find a rotational invariant form.  To see the property  let us consider a simple example.

 In Fourier transformation of standard plane wave we find that
$$ \delta(x)\cdot \delta(y)=\int {dk_x\over 2\pi} exp( i k_xx)~\int {dk_y\over 2\pi}exp( ik_yy) =\int {dk_x\over 2\pi} {dk_y\over 2\pi}exp( i k_xx+i k_yy)$$
$$=\int{dk_y\over 2\pi}\left[\int {dk_x\over 2\pi} exp( i x\left(k_x+{ k_yy\over x}\right)\right] = \int{dk_y\over 2\pi}\left[\delta(x)\right] =\delta(x)\cdot \delta(0).\eqno{(3.1)}$$
Thus, if we first shift the variable $k_x$ and take an integration over the shifted variable $k_x$ we find that $\delta(x)\cdot \delta(y)=\delta(x)\cdot \delta(0)$. In a same way, if we first shift the variable $k_y$ and take an integration over the shifted variable $k_y$ we find that $\delta(x)\cdot \delta(y)=\delta(0)\cdot \delta(y)$.

Now, to perform the Fourier transformation of plane wave (2.29) we can first shift the variable   $p_x$ then take an integration over the shifted variable $p_x$.  Next, we shift the variable $p_y$ and take an integration over the shifted variable $p_y$.  Finally, we  shift the variable   $p_z$ then take an integration over the shifted variable $p_z$.  After the lengthy calculation the result is
$$ \rho^R_h(r)\equiv \int{dp_x\over 2\pi}{dp_y\over 2\pi}{dp_z\over 2\pi} exp\left( i \vec x \cdot \vec p -{|\vec x|\over 4  h}(\vec x^2  \vec p^2 - (\vec x \cdot \vec p)^2)\right)\hspace{7cm}$$
$$\hspace{1cm}=\int{dp_x\over 2\pi}{dp_y\over 2\pi}{dp_z\over 2\pi} exp\left( i (p_xx+p_yy+p_zz) -{|R|\over 4  h}(R^2  (p_x^2+p_y^2+p_z^2) - (p_xx+p_yy+p_zz)^2)\right)$$
$$= {\delta (R)\over 2\pi h R}~e^{-{R\over h}{x^2+y^2\over z^2}}.\hspace{10.3cm}\eqno{(3.2)}$$
As that discussed in the Fourier transformation of standard plane wave,  when we  interchange the order of integrating variables $p_x$, $p_y$ and $p_z$ we can find that 
$$ \rho^R_h(r)={\delta (R)\over 2\pi h R}~e^{-{R\over h}{x^2+y^2\over z^2}}= {\delta (R)\over 2\pi h R}~e^{-{R\over h}{y^2+z^2\over x^2}}={\delta (R)\over 2\pi h R}~e^{-{R\over h}{z^2+x^2\over y^2}}.\eqno{(3.3)}$$
From above result we will in following discuss how to  find the smeared source form on fuzzy space.  

 First, the fuzzy space we have investigated is the one with  constant radius $X_x^2+X_y^2+X_z^2=R^2$.  Therefore, in our case the space is a fuzzy sphere with constant radius.  The fuzziness shall be shown on the fuzzy 2-sphere and not on the radius direction.  

Next, on the fuzzy sphere we know that 
$$R{x^2+y^2\over z^2}=R\tan\theta \approx R\theta=r, \eqno{(3.4)}$$
when ${x^2+y^2\over z^2}\ll 1$.  Note that $r$ is the distance from north pole on which the ``particle" source stands, as shown in figure 1. 
\\
\\

\scalebox{1}{\hspace{4cm}\includegraphics{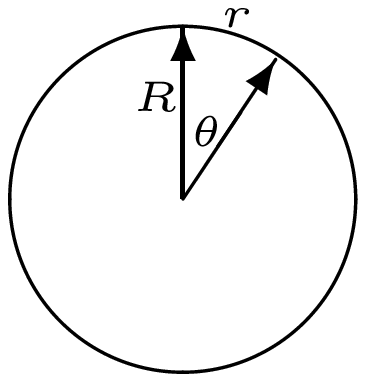}}
\\
{\it Figure 1: $r$ and $R$.  $R$ is the radius of fuzzy sphere and $r$ is the distance on fuzzy sphere.}
\\

Thus, in the case of $r\ll R$, the position which we consider  is very near the  source the  ``point source"  on the fuzzy 2-sphere will show a smeared distribution function described by
$$\rho_h(r)= C_2~ exp\left(-{r\over h}\right) = {1\over 2\pi h^2} exp\left(-{r\over h}\right), ~~~~~~2D~fuzzy~space,\eqno{(3.5)}$$
in which the $C_2$ is determined by the condition of $\int dx dy~ \rho_h(r) =1$.  

It is natural to assume that the exponential form $exp\left(-{r\over h}\right)$  is universal and thus on the fuzzy 3-sphere and fuzzy 4-sphere the point source will become a smeared one with distribution function described by
 $$\rho_h(r)= {1\over 8\pi h^3}~ exp\left(-{r\over h}\right), ~~~~~~3D~fuzzy~space\eqno{(3.6)}$$
$$\rho_h(r)= {1\over 12\pi^2 h^4}~ exp\left(-{r\over h}\right), ~~~~~~4D~fuzzy~space\eqno{(3.7)}$$
and general form
$$\rho_h(r)= {\Gamma(D/2)\over 2\pi^{n/2}\Gamma(D) h^n}~ exp\left(-{r\over h}\right), ~~~~~~D-dimentional~fuzzy~space.\eqno{(3.8)}$$
These are the main results calculated in this paper and it corresponds to the smeared source on Moyal space described in (1.10).  
\\

Let us make following comments to conclude this section.

1.  Note that in Moyal space as $[X,Y]=i\theta$ the function ${r^2\over \theta}$ is dimensionless and factor $exp\left(-{r^2\over \theta }\right)$ could appear in the Gaussian decay function of smeared source. Now,  on fuzzy space as $[X,Y]=ih Z $ the function ${r\over h}$ is dimensionless and factor $exp\left(-{r\over h}\right)$ could appear in the exponential decay function of smeared source.   Although the function ${r^2\over h^2}$ is also dimensionless, however, the Gaussian factor $exp\left(-{r^2\over h^2}\right)$ does not appear in the smeared source on fuzzy space.  This is our main finding in this paper that the fuzziness provides an  exponential decay smeared source rather then the Gaussian decay smeared source.  

2.  It shall be noticed that the  space on which the smeared described by (3.5) is  the 2D fuzzy sphere described by $X_1$,$X_2$,$X_3$ in  (2.1) and not the 3D space.  In the same way, it is natural to belive that the smeared function described by (3.7), for example, is the 4D sphere described by the condition
$$ \epsilon^{\mu\nu\lambda\sigma}X^\mu X_\nu X^\lambda X_\sigma= h X_\rho.\eqno{(3.9)}$$
and $$\sum_{\mu=1}^5~X^\mu X_\mu=R,\eqno{(3.10)}$$
where $R$ is the radius of the sphere [20].  It is a challenged problem to prove this conjecture.

3. As we are considering the case of a large radius $R$ and interesting the property at short distance (or near the source point) the geometry considered in following section could be regarded as a flat space with fuzziness.

4. Note also that the plane wave in (2.29) is that on the 3D space while the Fourier transformation of (3.5) shall give the plane wave on 2D fuzzy sphere.   Therefore we can use the smeared source functions (3.6)-(3.8) to perform the Fourier transformation and  find the plane wave on the fuzzy space.  The results are $\langle\vec p|\vec x\rangle =\Psi_{\vec p}(\vec x)$ with
$$\Psi_{\vec k}(\vec x) =~{1\over (1+h^2 k^2)^2}~exp\left(i \vec k\cdot \vec x\right),~~~~~~3D~fuzzy~space. \eqno{(3.11)} $$
$$\Psi_{\vec k}(\vec x) = {1-{1\over3}h^2k^2\over (1+h^2 k^2)^3}~exp\left(i \vec k\cdot \vec x\right),~~~~~~4D~fuzzy~space. \eqno{(3.12)} $$
For the general smeared function (3.8) we find that
$$\Psi_{\vec k}(\vec x) = {Im\left[(1+ihk)^{D-1}\right]\over (D-1) (1+h^2 k^2)^{D-1}hk}~exp\left(i \vec k\cdot \vec x\right),~~~~~~D-dimensional~ fuzzy~ space, \eqno{(3.13)} $$
which will be used in next section to study the thermodynamics  of gas on fuzzy space.

In next section we will see how the exponential decay function of smeared source will modify the UV behavior of quantum field and the thermodynamics of ideal gas.  We also compare it with that in the Gaussian decay function in Moyal space.
\section {Quantum Field and Thermodynamics on Fuzzy Space}
\subsection {Feynman Propagator on Fuzzy Space}
Using the smeared form we can now calculate the field propagator on fuzzy space in coherent state approach.  In D dimension  the Feynman propagator $G(k)$ is defined by 
$$G(x) =\int{dk^D\over (2\pi)^D}~ e^{i \vec k\cdot \vec x}~G(k), \eqno{(4.1)}$$
in which $G(x)$ satisfies the relation
$$ \nabla^2_x G(x) = -\rho_h(r). \eqno{(4.2)}$$
Thus 
$$\int{dk^D\over (2\pi)^D} (k^2)~e^{i \vec k\cdot \vec x} G(k)= \rho_h(r), \eqno{(4.3)}$$
and
$$G(k)={1\over k^2}~ \int dx^D~e^{i \vec k\cdot \vec x}~\rho_h(r).\eqno{(4.4)}$$
Substituting the smeared function (3.6) and (3.7) into (4.4) we find that 
$$G(k) = {1\over k^2}~{1\over (1+h^2 k^2)^2},~~~~~~3D~fuzzy~space. \eqno{(4.5)} $$
$$G(k) = {1\over k^2}~{1-{1\over3}h^2k^2\over (1+h^2 k^2)^3},~~~~~~4D~fuzzy~space. \eqno{(4.6)} $$
Thus, in the UV region ($k\gg 1$) the fuzzy space could provide an extra power decay factor $(hk)^{-4}$ in the propagator, contrast to the exponential decay factor $e^{-\theta k^2}$  in the Moyal space.  

For the general smeared function (3.8) we find that
$$G(k) = {1\over k^2}~{Im\left[(1+ihk)^{D-1}\right]\over (D-1) (1+h^2 k^2)^{D-1}hk},~~~~~~D-dimensional~ fuzzy~ space, \eqno{(4.7)} $$
in which $Im[g]$ means the image part of $g$. In the limit of large $k$ we find that 
$$G(k) \approx\cases{\sim~{1\over k^2}~{1\over (hk)^D} ,&~~~even~D\cr
	\sim~{1\over k^2}~~{1\over (hk)^{D+1}},&~~~odd~~D,\cr}\eqno{(4.8)} $$
which shows that, in the UV region ($k\gg 1$) the D-dimensional fuzzy space could provide an extra power decay factor in the propagator.  Note that the  extra decay factor could render the quantum field to be a finite theory without any renormalization as a simple power counting shows that the internal line gives $\int d^Dk~ G(k)$ which does not have a UV divergent contribution.
\subsection {Thermodynamics on Fuzzy Space}
We now investigate the thermodynamics of non-relativistic  ideal gases on the noncommutative fuzzy geometry.  To begin with let us recall the  thermodynamics of ideal gases on the Moyal geometry in coherent state approach [21].  

First, the key principle of statistical mechanics is that the probability of a point gas having energy $E$ is proportional to $e^{-\beta E}$ [22].  Therefore we shall consider the value
$$  <X|e^{-\beta H}|X> = \sum_p <X|p><p|e^{-\beta H}|p><p|X>$$
$$= {1\over (2\pi)^3} \int d^3k~   e^{-{ \vec k^2\over 2mT} -{\theta \vec k^2\over 2}} = {1\over (2\pi)^3} \int d^3k~   e^{-{\vec k^2\over 2mT}(1+m\theta T)},\eqno{(4.9)}$$
as the plane wave in Moyal space is $exp(ik\cdot y -{\theta \vec k^2\over 4})$.  Thus the associated Boltzmann factor becomes
$$P(k^2)\sim e^{-{\vec k^2\over 2mT}(1+m\theta T)}.\eqno{(4.10)}$$
This means that use  the following simple substitution
$$ T ~~~~~ \Rightarrow ~~~~~T_\theta \equiv  {T\over 1+m\theta T}, \eqno{(4.11)}$$
we could obtain the desired thermal property on the Moyal background form that in commutative case.

  For example, the  energy $U_\theta(T)$ of gas on Moyal  space  at temperature $T$ is equal to the  energy $U(T_\theta)$ of gas on commutative space at temperature $T_\theta$, i.e. $U_\theta(T)=U(T_\theta)$.   Thus the heat capacity $C_\theta(T)$ of ideal gas at temperature $T$ on Moyal space is 
$$C_\theta(T)\equiv {dU_\theta(T)\over dT}={dT_\theta\over dT}{dU_\theta(T)\over dT_\theta}={dT_\theta\over dT}{dU(T_\theta)\over dT_\theta}= {1\over (1+m\theta T)^2}C(T_\theta).\eqno{(4.12)}$$
As the heat capacity, $C(T_\theta)$, of non-relativistic gas on commutative space is a constant value  [23] above relation implies that heat capacity of the non-relativistic gas on Moyal space will approach to zero in high temperature limit.  The property also shows in fuzzy space as will be analyzed in below.

Let us now turn to the case of the non-relativistic gas on 3D fuzzy space.  Then, using the plane wave (3.11) and the formula (4.9) we see that the Boltzmann factor becomes
$$P(k^2)\sim {e^{-k^2\over2mT}\over (1+h^2k^2)^4}.\eqno{(4.13)}$$
Through standard calculations we find that the  heat capacity of non-relativistic gas $C(T)$ is
$$C_h(T)=-{1\over 384  h^{13}  m^5 T^5}\left[\sqrt{2\pi\over h^2 mT}\left(h^2 mT  +14(h^2 mT)^2 +33(h^2 mT)^3\right)\right.\hspace{3cm}$$
$$\left.+\pi\left(1+15h^2 mT  +45(h^2 mT)^2 +15(h^2 mT)^3\right)e^{1\over 2h^2mT}~(Erf(\sqrt{{1\over 2h^2mT}})-1)\right].\eqno{(4.14)}$$
 In the low and high temperature 
we find that 
$$C_h(T) \approx\cases{{15\over 128h^3}\sqrt{2\pi h^2mT}~\left(21+100h^2 mT \right),& low~temperature,\cr
	{5\pi\over 128h^3h^7m^2T^2},& high~temperature,\cr}\eqno{(4.15)} $$
and heat capacity approaches to  zero in the low and high temperature limits, as shown in figure 2 after analytic evaluation. 
\
\\

\scalebox{1}{\hspace{4cm}\includegraphics{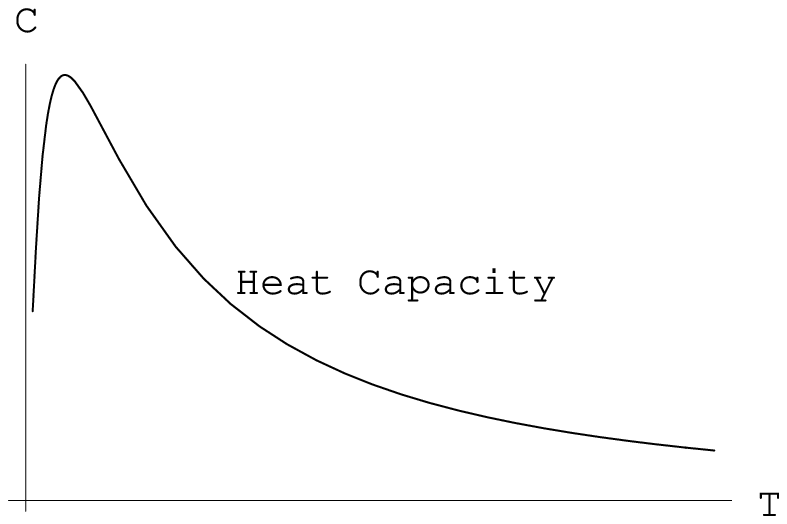}}
\\
{\it Figure 2: Heat capacity of non-relativistic gas on fuzzy space.  The heat capacity of non-relativistic gas on fuzzy space approaches to zero in the low and high temperature limits.}
\\

Let us make following comments to conclude this section.

1.  The capacity $C_\theta(T)$ in (4.12) shows a  non-perturbative property in high temperature while the capacity $C_h(T)$ in (4.14) shows a  non-perturbative property in both low and high temperatures.

2. In [24] we had studied the thermodynamics on fuzzy space in conventional approach.   It also finds that the heat capacity of non-relativistic gas on fuzzy space approaches to  zero in the low and high temperature limits.  In that paper we consider, for example, the gas on fuzzy 2-sphere with radius $R$.   As the system has the spectrum $E= {\ell(\ell+1)\over 2mR^2}$ with $\ell=0 \cdot \cdot \cdot J$ it could have a maximum energy level, which reveals the fuzzy property of space. Thus the system has bounded value of energy and the  heat capacity shall become zero  in high temperature.  In this paper we study the thermodynamics on fuzzy space in coherent state approach.  The spectrum does not be bounded, as $-\infty < k< \infty$, and the zero value of heat capacity  in high temperature reveals the smeared source property of fuzzy space.

\section {Black Hole on Fuzzy Space}
In this section we will see how the exponential decay function of smeared source will modify black hole properties and compare it with that in the Gaussian decay function in Moyal space.
\subsection {Fuzzy Schwarzschild Black Hole}
We can now use the smeared source (3.7) to find the 4D Schwarzschild solution following the method of  Nicolini [9].  First, the temporal component of the energy momentum tensor $T^\nu_\mu$ is identified as, $T^t_t=-\rho_h$.   Next, demanding the condition on the metric coefficients $g_{tt} = -(g_{rr})^{-1}$ for the noncommutative Schwarzschild-like metric and using the covariant conservation of energy momentum tensor, the energy momentum tensor can be fixed to the form [9],
$$T^\nu_\mu = diag[-\rho_h ,p_r,p_\perp,p_\perp], \eqno{(5.1)} $$
in which 
$$p_r= -\rho_h, ~~~~~~~~p_\perp=p_r-{r\over2}\partial_r \rho_h.\eqno{(5.2)}$$
Thus, rather than a massive, structureless point, a source turns out to a self-gravitating, droplet of anisotropic fluid of  density $\rho_h(r)$, radial pressure $p_r$  and tangential pressure $p_\perp$.  On physical grounds, a non-vanishing radial pressure is needed to balance the inward gravitational pull, preventing 
droplet to collapse into a matter point. This is the basic physical effect on matter caused by spacetime noncommutativity or fuzziness [9]. 

The solution of Einstein equation, using (5.1) as the matter source, is  the same as replacing the mass of  Dirac-delta function source in standard Schwarzschild spacetime by the effective mass of  smeared source, i.e.  
$$ M=M \int d^3 x \delta^3(\vec x)~~~\rightarrow~~~M_h(r) = M \int d^3 x  \rho_h(r) = M\left[1- \left({1\over2}\left({r\over h}\right)^2+{r\over h}+1\right)e^{-r/h}\right],\eqno{(5.3)}$$
which approaches to  $M$ in the limit $h \rightarrow 0$, the commutative space.
The geometry of fuzzy black hole is thus described by the line element
$$ds^2 = -\left(1-{2M_h(r)\over r} \right)dt^2+\left(1-{2M_h(r)\over r} \right)^{-1}dr^2 +r^2 \left(d\theta^2+\sin\theta^2 d\phi^2\right).  \eqno{(5.4)}$$
Note that the effective mass described in Moyal space with smeared source of (1.10) is $M_\theta(r) ={2M\over \sqrt \pi}\gamma(3/2,r^2/4\theta)$ (where $\gamma(3/2,r^2/4\theta)$ is the lower incomplete Gamma function [9]).  It approaches to  $M$ in the limit $\theta\rightarrow 0$, the commutative space.  

To discuss the black hole property we see that the solution of condition $g_{00}=0$ i.e.
$$ r_H=2M \left[1- \left({1\over2}\left({r\over h}\right)^2+{r\over h}+1\right)e^{-r/h}\right], \eqno{(5.5)}$$ 
will give the radius of horizon $r_H$.  We plot in figure 3 the mass versus  horizon solution.  We see that there are two horizons:  cosmological horizon $r_C$ and event horizon $r_H$ with $r_C <r_H$, as that in the noncommutative case [9]. We also see that the minimum mass $M = M_c\approx 2.56$  occurs at $r_C =r_H\equiv r_*\approx 3.4$. 
\\
\\

\scalebox{1}{\hspace{2cm}\includegraphics{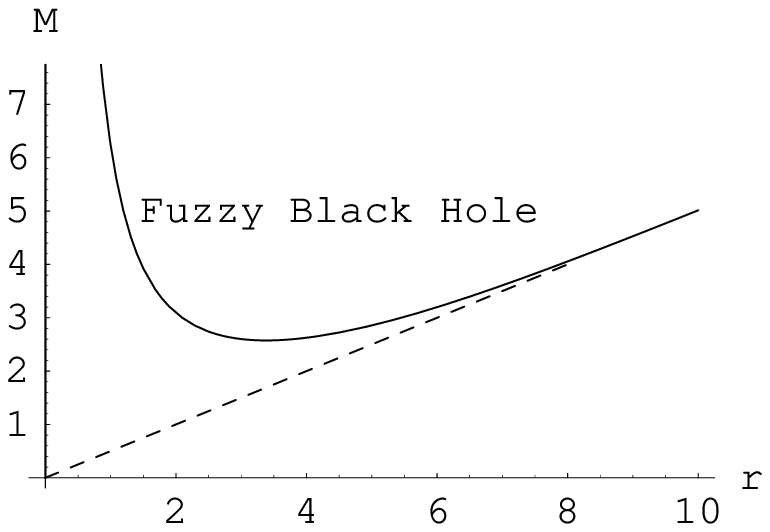}}
\\
{\it Figure 3: Mass versus cosmological horizon $r_C$ and event horizon $r_H$. 
Dashing line describes the standard Schwarzschild black hole.  Solid line describes the fuzzy black hole which has a minimum mass $M = M_c\approx 2.56$ at $r_C =r_H\equiv r_*\approx 3.4$.}
\\

As the spacetime does not have horizon once the black hole mass is less then a critical mass $M_c$ a large black hole will therefore stop to evaporate after  its mass is less $M_c$.   To see the property we then analyze the black hole temperature in below. 

  The horizon radius $r_H$ read from $g_{00}$ is defined by
$$r_H = 2M_h(r_H)=2M\left[1- \left({1\over2}\left({r_H\over h}\right)^2+{r_H\over h}+1\right)e^{-r_H/h}\right],\eqno{(5.6)}$$
which reduces to $r_H = 2M$ in the large $M$ limit.  Thus the noncommutativity in the fuzzy space could be shown only in the small black hole case.   Using the above relation the Hawking temperature is calculated by
$$T_H\equiv \left({1\over 4\pi}  {d g_{00}\over dr}\right)_{r=r_H} = {1\over 4\pi r_H}\left[1-{1\over2}\left(r_H\over h\right)^3{e^{-r_H/h}\over 1- \left({1\over2}\left({r_H\over h}\right)^2+{r_H\over h}+1\right)e^{-r_H/h}}\right].  \eqno{(5.7)}$$
which reduces to $T_H = {1\over 4\pi r_H}$ in the large $r_H$ limit.  Using above formula we plot in figure 4 the black hole temperature $T_H $ vs $r_H$. 
\\

 \scalebox{1}{\hspace{2cm}\includegraphics{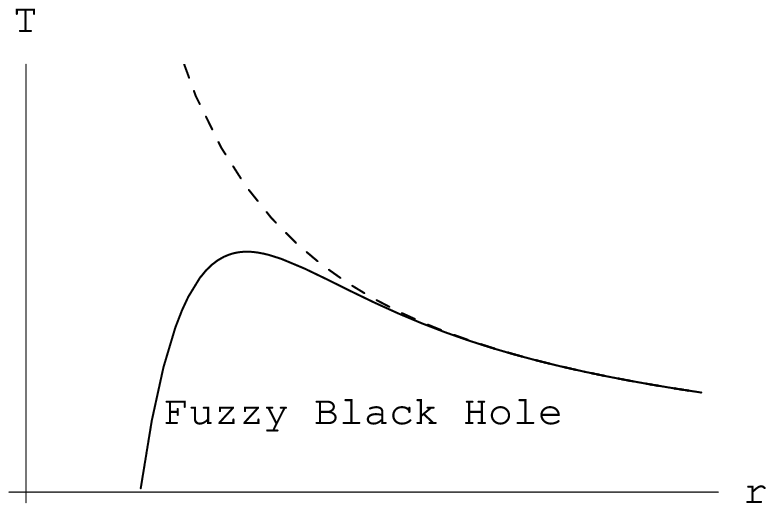}}
\\
\\
{\it Figure 4: Black hole temperature $T_H $ vs horizon radius $r_H$. Dashing line describes the standard Schwarzschild black hole.  Solid line describes the fuzzy black hole which has a maximum temperature.}
\\
\\
We see that a fuzzy black hole has a maximum temperature and $T_H  =0 $ at a finite value of $r_H =r_*\approx 3.4$, which corresponds to a stable remnant with mass  $M=M_c\approx 2.56$ in figure 3.   These properties are the same as those in the noncommutative case [9].
\subsection {Fuzzy BTZ Black Hole}
The noncommutative non-rotating BTZ black hole had been investigated in [12].  They found that the smeared Gaussian distribution ($\rho\sim e^{-r^2/\theta}$) around the origin is not appropriate to make a small black hole and there exist uncertainty for the thermal quantities. Therefore they study the problem with the non-Gaussian type of smeared Maxwell distribution source ($\rho\sim r e^{-r^2/\theta}$). In this case the  small black hole have a well defined property.   While the Maxwell distribution source  does not have a really physical motivation it is interesting to see whether non-Gaussian type of smeared distribution source  ($\rho\sim e^{-r/h}$) found in the fuzzy space could have a well defined  thermal quantities  for a small black hole.

Let us consider the general 1+2D matter density 
$$ \rho_h(r) = {Mr^n\over 2^{n+1}\pi h^{2+n}}~ exp\left(-{r\over h}\right),~~~n=0,1.  \eqno{(5.8)}$$
Case of $n=0$ is that on the fuzzy space. On other hand, the case of  $n=1$ is used to compare the  the Maxwell distribution source in noncommutative case [12].  The static metric ansatz of non-rotating BTZ black hole is described by [12]
$$ds^2= -g_{00}dt^2 + g_{00}^{-1}dr^2 + r^2d\phi^2.  \eqno{(5.9)}$$
As in Schwarzschild black hole we consider matter's energy-momentum
$$T^\nu_\mu = diag[-\rho_h ,p_r,p_\phi], \eqno{(5.10)} $$
and the Einstein equation implies that
$$g_{00}={r^2\over\ell^2}- 16 \pi G  \int_0^r dr r \rho_h(r) = \cases{{r^2\over\ell^2}- 8 G M\left[1- \left({r\over h}+1\right)e^{-r/h}\right] ,&n=0\cr
	{r^2\over\ell^2}- 8 G M\left[1- \left({1\over2}\left({r\over h}\right)^2+{r\over h}+1\right)e^{-r/h}\right],&n=1.\cr}\eqno{(5.11)}$$
and 
$$p_r= -\rho_h, ~~~~~~~~p_\phi=(rp_r)'.\eqno{(5.12)}$$
To discuss the BTZ black hole property we see that the solution of condition $g_{00}=0$ i.e.
$$ {8GM\over h^2}=\cases{{r_H^2\over h^2}\left[1- \left({r_H\over h}+1\right)e^{-r_H/h}\right]^{-1}\approx 2+{4\over3}{r_H\over h} +O(r_H^2),&n=0\cr
	{r_H^2\over h^2}\left[1- \left({1\over2}\left({r_H\over h}\right)^2+{r_H\over h}+1\right)e^{-r_H/h}\right]^{-1}\approx {6h\over r_H}+{9\over 2} +O(r_H), &n=1.\cr}, \eqno{(5.13)}$$ 
will give the radius of horizon.  We plot in figure 5 the mass versus  horizon solution. We see that case of $n=1$ could have two horizons which converge to a single degenerate horizon at $r_{min}$.   However, the case of $n=0$ has only one horizon : the event horizon which become zero radius at finite mass $M={h^2\over 4G}$, as calculated in (5.13).
\\
\\

\scalebox{1}{\hspace{2cm}\includegraphics{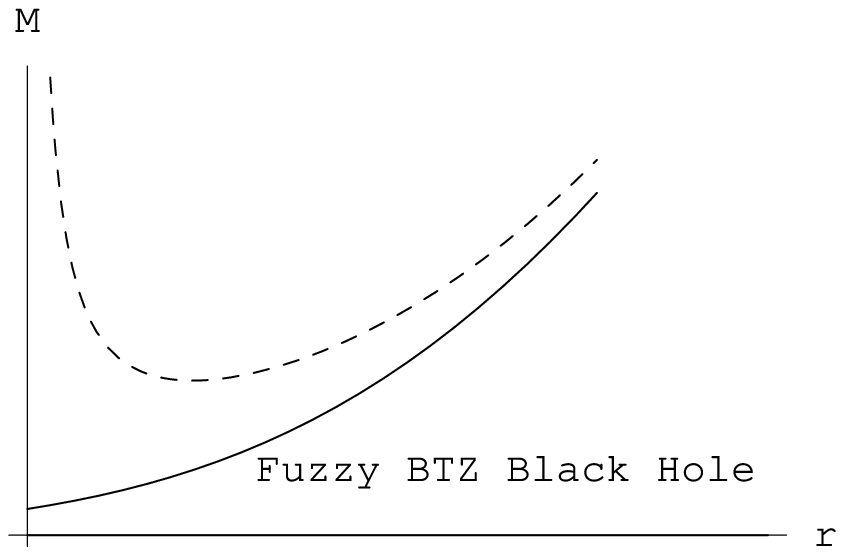}}
\\
{\it Figure 5: Mass versus horizon $r$. Solid line describes the fuzzy BTZ black hole which has only one horizon : the event horizon which become zero at finite mass.  Dashed line describes the case of $n=1$ that could have two horizons which converge to a single degenerate horizon at $r_{min}$.}
\\

Using the above relation the Hawking temperature is calculated by
$$T_H\equiv \left({1\over 4\pi}  {d g_{00}\over dr}\right)_{r=r_H} = \cases{{r_H\over 4\pi\ell^2}\left[2- {r^2\over h^2}\left(e^{r_H/h}-\left({r_H\over h}+1\right)\right)^{-1}\right] ,&n=0\cr
	{r_H\over 4\pi\ell^2}\left[2- {r_H^2\over h^2}\left(e^{r_H/h}-({1\over2}\left({r_H\over h}\right)^2+{r_H\over h}+1)\right)^{-1}\right],&n=1.\cr}.  \eqno{(5.14)}$$
The black hole entropy $S$ can be calculated from the first-law of thermodynamics:
$$S =\int {dM\over T_H} =\int^{r_H} {1\over T_H}{dM\over dr_H}dr_H = 
\cases{{\pi^2\over G}\int_0^{r_H} {dr_H\over 1-\left({r_H\over h}+1\right)e^{-r_H/h}}, & n=0\cr
{\pi^2\over G}\int_{r_{min}}^{r_H} {dr_H\over 1-\left({1\over2}\left({r_H\over h}\right)^2+{r_H\over h}+1\right)e^{-r_H/h}}, & n=1\cr}. \eqno{(5.15)}$$
For the case of $n=0$, which is the fuzzy BTZ black hole, we have to integrate the radius $r_H$ from zero, as the minimum radius of fuzzy BTZ black hole is zero.  However, this integration is infinite and we could not define the thermal quantities.  The case of $n=1$ does not suffer such a problem as the minimum radius is a finite value $r=r_{min}$ (as shown in figure 5).

In conclusion, although the fuzzy space could provide a non-Gaussian type of smeared source it could  not  have a well defined thermal quantities  for a small  BTZ black hole. The properties are the same as those in the noncommutative case [12].   It feels short for the author's expectation and the criterion for existence a regular black hole with a general smeared source is given in next section.

\section {Criterion for Existence a Regular Black Hole}
\subsection {Regular BTZ Black Hole}
Consider first the BTZ black hole with general density $\rho_h = M f(r)$. Then  
$$g_{00}={r^2\over\ell^2}- 16 \pi G  \int_0^r dr r \rho_h(r) ={r^2\over\ell^2}- 16 \pi GM  \int_0^r dr r f(r). \eqno{(6.1)}$$
Thus the relation of black mass $M$ and its horizon radius $r_H$ is
$$ M ={r_H^2\over\ell^2}{1\over 16 \pi G  \int_0^{r_H} dr r f(r)}. \eqno{(6.2)}$$
If the density function at small radius behaves as 
$$ \rho_h = M f(r) \approx M c r^n, \eqno{(6.3)}$$
then (6.2) implies that the black hole mass at small $r_H$  is
$$ M \approx {n+2\over 16 \pi G c \ell^2}{1\over r^n} . \eqno{(6.4)}$$
Therefore, only if $n>0$ the black hole mass at small $r_H$ radius could become infinite.  In this case we can from figure 3 see that the black hole has two horizons and it will reach a finite maximum temperature before cooling down to absolute zero. Thus it leaves a stable remnant and we have a regular black hole.
\subsection {Regular Kaluza-Klein Black Hole}
Consider next the D-dimensional Schwarzschild black hole with general density $\rho_h = M f(r)$. Then  [13]
$$g_{00}=1- {2GM\over r^{D-3}}\Omega_{D-2} \int_0^r dr r^{D-2} f(r). \eqno{(6.5)}$$
Thus the relation of black mass $M$ and its horizon radius $r_H$ is
$$ M ={r_H^{D-3}\over\Omega_{D-2} \int_0^r dr r^{D-2} f(r)}. \eqno{(6.6)}$$
If the density function at small radius behave as (6.3) then (6.6) implies that the black hole mass at small $r_H$  is
$$ M \approx {D+n-1\over 2 G c \Omega_{D-2}}{1\over r^{n+2}} . \eqno{(6.7)}$$
Therefore, only if $n>-2$ the black hole mass at small $r_H$ radius could become infinite.  In this case we can from figure 1 see that the black hole has two horizons and it will reach a finite maximum temperature before cooling down to absolute zero. Thus it leaves a stable remnant and we have a regular black hole.   
   
Above criterion is consistent with the all known results [12,13].
\section {Discussion}
In this paper we first use the coherent state formalism of fuzzy space to show that the  fuzziness  will eliminate point-like structure of a particle in favor of smeared object, which is an exponential decay function in contrast to the Gaussian type in the Moyal noncommutative space.   We also see that the exponential decay function could, in the UV region of 4D fuzzy space, provides propagator with an extra power decay factor $(hk)^{-4}$, contrast to the exponential decay factor $e^{-\theta k^2}$  in the Moyal space.   Both decay function could render the quantum field to be a finite theory without any renormalization, just after a simple power counting.  We also see that the heat capacity of  non-relativistic gas will  approaches to zero at high temperature.

 Using the behavior of smeared  source we have studied the fuzzy 4D Schwarzschild-like geometry and see that the fuzzy black hole can reach a finite maximum temperature before cooling down to absolute zero and leave a stable remnant, as that in the noncommutative case.  We next study the properties of fuzzy 3D BTZ and find that although the fuzzy space could provide a non-Gaussian type of smeared source it could  not  have a well defined thermal quantities  for a small  BTZ black hole. The properties are the same as those in the noncommutative case.  Finally, We have presented a useful criterion for existence a regular BTZ black hole or  regular Kaluza-Klein black hole with a general smeared source.   
\\
\\
\begin{center} {\bf REFERENCES}\end{center}
\begin{enumerate}
\item H. S. Snyder, Phys. Rev. 71 (1947) 38; 72 (1947) 68.
\item Connes, A. Connes, ``Noncommutative Geometry", Academic Press, New York, 1994.;\\ Connes, ``A.Gravity coupled with matter and the foundation of non commutative geometry, Comm. in Math. Phys. 182 155-177 (1996), hep-th/9603053 .
\item   M. R. Douglas and N. A. Nekrasov, ``Noncommutative Field Theory ",  Rev.Mod.Phys.73 (2001) 977 [hep-th/0106048];\\
R. J. Szabo, Quantum Field Theory on Noncommutative Spaces, Phys. Rept. 378 (2003) 207 [hep-th/0109162]; A. P. Balachandran, S. Kurkcuoglu and S. Vaidya, ``Lectures on Fuzzy and Fuzzy SUSY Physics ", [hep-th/0511114].
\item  A. Connes, M. R. Douglas and A. Schwarz, ``Noncommutative Geometry and Matrix Theory: Compactification on Tori", JHEP 9802:003 (1998) [hep-th/9711162]. 
\item  N. Seiberg and E. Witten, ``String Theory and Noncommutative Geometry", JHEP 9909 (1999) 032 [hep-th/9908142]. 
\item Wung-Hong Huang, "Effective Potential of Noncommutative Scalar Field Theory: Reduction of Degree of Freedom by Noncommutativity", Phys.Rev. D63 (2001) 125004  [hep-th/0101040]; \\ A. V. Strelchenko, D. V. Vassilevich, ``On space-time noncommutative theories at finite temperature ", Phys.Rev.D76 (2007) 065014; [arXiv:0705.4294[hep-th]]; \\ C. D. Fosco, G. A. Silva,``Noncommutative real scalar field theory in 2+1 dimensions at finite temperature ", JHEP0801 (2008) 011 [arXiv:0710.3836v2 [hep-th]].
\item  S. Cho, R. Hinterding, J. Madore, H. Steinacker,``Finite Field Theory on Noncommutative Geometries," Int.J.Mod.Phys. D9 (2000) 161 [hep-th/9903239];\\
A. Smailagic and E. Spallucci,``Feynman Path Integral on the Noncommutative Plane, " J.Phys. A36 (2003) L467 [hep-th/0307217 ]; ``UV divergence-free QFT on noncommutative plane,"  J. Phys. A36 (2003) L517 [hep-th/0308193 ]. 
\item  A. Smailagic and E. Spallucci,``Lorentz invariance and unitarity in UV finite NCQFT ,"  J.Phys. A37 (2004) 1 [hep-th/0406174];\\A. Gruppuso,``Newton's law in an effective non commutative space-time ," J.Phys. A38 (2005) 2039 [hep-th/0502144 ].
\item  P. Nicolini,`` A model of radiating black hole in noncommutative geometry ," J.Phys. A38 (2005) L631 [hep-th/0507266];\\ P. Nicolini, A. Smailagic, and E. Spallucci ,``Noncommutative geometry inspired Schwarzschild black hole," Phys.Lett. B632 (2006) 547 [gr-qc/0510112].
\item  S. Ansoldi, P. Nicolini, A. Smailagic, and E.Spallucci ,``Noncommutative geometry inspired charged black holes ," Phys.Lett. B645 (2007) 261 [gr-qc/0612035];\\E. Spallucci, A. Smailagic and P. Nicolini,``Non-commutative geometry inspired higher-dimensional charged black holes,''  Phys.  Lett. B670 (2009) 449   [arXiv:0801.3519];\\P, Nicolini,``Noncommutative Black Holes, The Final Appeal To Quantum Gravity: A Review," Int. J. Mod. Phys. A 24 (2009) 1229-1308 [arXiv:0807.1939 ];\\ Nicolini and E. Spallucci, `Noncommutative geometry inspired wormholes and dirty black holes,''  Class. Quant. Grav. 27 (2010) 015010 [arXiv:0902.4654]. 
\item  Y. S. Myung, Y.-W. Kim, and Y.-J. Park ,``Thermodynamics and evaporation of the noncommutative black hole ,"  JHEP 0702 (2007) 012 [gr-qc/0611130];\\ R. Banerjee, B. R. Majhi, S. Samanta,`` Noncommutative Black Hole Thermodynamics," Phys.Rev.D77 (2008) 124035 [arXiv:0801.3583];\\  R. Banerjee, B. R. Majhi, S. K. Modak,``Noncommutative Schwarzschild Black Hole and Area Law ,"  Class.Quant.Grav.26 (2009) 085010 [arXiv:0802.2176].
\item  Y. S. Myung and M. Yoon,``Regular black hole in three dimensions," Eur.Phys.J.C62 (2009) 405 [0810.0078];\\Mu-in Park,``Smeared Hairs and Black Holes in Three-Dimensional de Sitter Spacetime," Phys.Rev.D80 (2009) 084026 [arXiv:0811.2685].   
\item  K.Nozari and S. H. Mehdipour,``Parikh-Wilczek Tunneling from Noncommutative Higher Dimensional Black Holes," JHEP 0903 (2009) 061 [arXiv:0902.1945];\\
 T. G. Rizzo,``Noncommutative Inspired Black Holes in Extra Dimensions," JHEP 0609 (2006) 021 [hep-ph/0606051].
\item R. C. Myers, ``Dielectric-branes," JHEP 9912 (1999) 022 [hep-th/9910053];\\
C. Bachas, M. Douglas and C. Schweigert, ``Flux stabilization of D-branes,"  JHEP 0005 (2000) 048 [hep-th/0003037];\\ A. Y. Alekseev, A. Recknagel and V. Schomerus, ``Non-commutative world-volume geometries: Branes on su(2) and fuzzy sphere" JHEP 9909 (1999) 023 [hep-th/9908040].
\item C. Bachas, M. Douglas and C. Schweigert, ``Brane dynamics in background  and non-commutative geometry," JHEP 0005 (2000) 010 [hep-th/0003187]; \\ Pei-Ming Ho, ``Fuzzy Sphere from Matrix Model" JHEP 0012 (2000) 015 [hep-th/0010165]; \\K. Hashimoto and K. Krasnov, ``D-brane Solutions in Non-Commutative Gauge Theory on Fuzzy Sphere," Phys. Rev. D64(2001) 046007 [hep-th/0101145];\\ S. Iso, Y. Kimura, K. Tanaka, and K. Wakatsuki, "Noncommutative Gauge Theory on Fuzzy Sphere from Matrix Model," Nucl. Phys. 604(2001) 121 [hep-th/0101102].  
\item   H. Grosse, C. Klimcik and P. Presnajder, ``Towards finite quantum field theory in noncommutative geometry," Int. J. Theor. Phys. 35 (1996) 231  [hep-th/9505175]; ``On Finite 4D Quantum Field Theory in Non-Commutative Geometry", Commun.Math.Phys. 180 (1996) 429 [hep-th/9602115];\\
J. Gomis, T. Mehen and M.B. Wise, ``Quantum Field Theories with Compact Noncommutative Extra Dimensions," JHEP 0008, 029 (2000) [hep-th/0006160]. 
\item  S. Vaidya, ``Perturbative dynamics on fuzzy $S^2$ and $RP^2$", Phys.Lett. B512 (2001) 403 [hep-th/0102212]; \\C. S. Chu, J. Madore and H. Steinacker, ``Scaling Limits of the fuzzy sphere at one loop", JHEP 0108 (2001) 038  [hep-th/0106205];\\
B. P. Dolan, D. O. Connor and P. Presnajder, ``Matix $\phi^4$ models on the fuzzy sphere and their continuum limits", JHEP 0203 (2002) 013 [hep-th/0109084].
 \item Wung-Hong Huang, "Effective Potential on Fuzzy Sphere", JHEP 0207 (2002) 064 [hep-th/0203051]; ``Quantum Stabilization of Compact Space by Extra Fuzzy Space", Phys.Lett. B537 (2002) 311 [hep-th/0203176]; ``Casimir Effect on the Radius Stabilization of the Noncommutative Torus", Phys.Lett. B497 (2001) 317 [ hep-th/0010160]; ``Finite-Temperature Casimir Effect on the Radius Stabilization of Noncommutative Torus ", JHEP 0011 (2000) 041 [hep-th/0011037].   
 \item  J. R. Klauder and B. Skagerstam,``Coherent states", World
Scientific, Singapore, 1985;\\ J. M. Radcliffe. J. Phys. A 4 (1971) 313;\\ F. T. Arecchi, E. Courtens, R. Gilmore, and H. Thomas. Phys. Rev. A6 (1972) 2211.
\item  J. Castelino, S. Lee and W. Taylor IV, `` Longitudinal 5-branes as 4-shpere in Matrix Theory," Nucl. Phys. B526  (1998)  334 [hep-th/9712105]; \\S. Ramgoolam,`` On spherical harmonics for fuzzy spheres in diverse dimensions," Nucl.Phys. B610 (2001) 461 [hep-th/0105006]; ``Higher dimensional geometries related to fuzzy odd-dimensional spheres,"  JHEP 0210 (2002) 064 [hep-th/0207111].
\item   Wung-Hong Huang and Kuo-Wei Huang, "Thermodynamics on Noncommutative Geometry in Coherent State Formalism", Phys. Lett. B670 (2009) 416 [arXiv:hep-th/0808.0324]; 
\item  R. P. Feynman, Statistical Mechanics (1972).
\item  R. K. Pathria, Statistical Mechanics (Pergamon, London, 1972).
\item Wung-Hong Huang, "Thermodynamics on Fuzzy Space", JHEP 0908 (2009) 102 [arXiv:0901.0614].

\end{enumerate}
\end{document}